\documentclass[%
 reprint,
superscriptaddress,
 amsmath,amssymb,
 aps,pre,hyphens,floatfix
]{revtex4-2}
\usepackage{amsfonts}
\usepackage{amsmath}
\usepackage[]{amsmath}
\usepackage{amssymb}
\usepackage{multirow} 
\usepackage{enumerate}
\usepackage[usenames,dvipsnames,svgnames,table]{xcolor}
\usepackage{IEEEtrantools}
\usepackage{graphicx}
\usepackage[normalem]{ulem}  

\begin{document}

\preprint{APS/123-QED}

\title{The tricritical point of tricritical directed percolation is determined 
\\based on neural network
}

\author{Feng Gao}
\affiliation{Key Laboratory of Quark and Lepton Physics (MOE) and Institute of Particle Physics, Central China Normal University, Wuhan 430079, China}

\author{Jianmin Shen}
\affiliation{College of engineering and technology, Baoshan University, Baoshan 678000, China}
\affiliation{Key Laboratory of Quark and Lepton Physics (MOE) and Institute of Particle Physics, Central China Normal University, Wuhan 430079, China}

\author{Shanshan Wang}
\affiliation{Key Laboratory of Quark and Lepton Physics (MOE) and Institute of Particle Physics, Central China Normal University, Wuhan 430079, China}

\author{Wei Li}
\email[]{liw@mail.ccnu.edu.cn}
\affiliation{Key Laboratory of Quark and Lepton Physics (MOE) and Institute of Particle Physics, Central China Normal University, Wuhan 430079, China}
\affiliation{École Supérieure d'Informatique Électronique Automatique, Ivry-sur-Seine 94200, France}

\author{Dian Xu}
\affiliation{Key Laboratory of Quark and Lepton Physics (MOE) and Institute of Particle Physics, Central China Normal University, Wuhan 430079, China}
\date{\today}

\begin{abstract}
In recent years, neural networks have increasingly been employed to identify critical points of phase transitions. For the tricritical directed percolation model, its steady-state configurations encompass both first-order and second-order phase transitions. Due to the presence of crossover effects, identifying the critical points of phase transitions becomes challenging. This study utilizes Monte Carlo simulations to obtain steady-state configurations under different probabilities $p$ and $q$, and by calculating the increments in average particle density, we observe first-order transitions, second-order transitions, and regions where both types of transitions interact.These Monte Carlo-generated steady-state configurations are used as input to construct and train a convolutional neural network, from which we determine the critical points $p_{c}$ for different probabilities $q$. Furthermore, by learning the steady-state configurations associated with the superheated point $p=p_u$, we locate the tricritical point at $q_{t}=0.893$. Simultaneously, we employed a three-output CNN model to obtain the phase transition boundaries and the range of the crossover regions. Our method offers a neural network-based approach to capture critical points and distinguish phase transition boundaries, providing a novel solution to this problem.

\end{abstract}
\maketitle

\section{Introduction}
\label{intro}

Phase transitions and critical phenomena have received considerable attention in the field of statistical physics~\cite{baus2008equilibrium,hinrichsen2006non,hinrichsen2000non,domb2000phase}. While much is understood about equilibrium phase transitions~\cite{baus2008equilibrium}, our knowledge of non-equilibrium phase transitions remains limited~\cite{lubeck2004universal,dalla2010quantum,menyhard1995non}. Among the most extensively studied models in this area is the contact process (CP)\cite{harris1974contact,windus2007phase,windus2008cluster,da2011critical}, which captures the dynamics of particle transfer and annihilation on a lattice. The CP model has significant real-world applications, ranging from the spread of infectious diseases\cite{mossong2008social} to species reproduction~\cite{sankaran2019clustering} and various reaction-diffusion phenomena involving particles~\cite{schutz1995reaction,pastor2015epidemic}.

\begin{figure*}[!thb]
\centering
\begin{tabular}{ccc}   
    \includegraphics[width=0.32\textwidth]{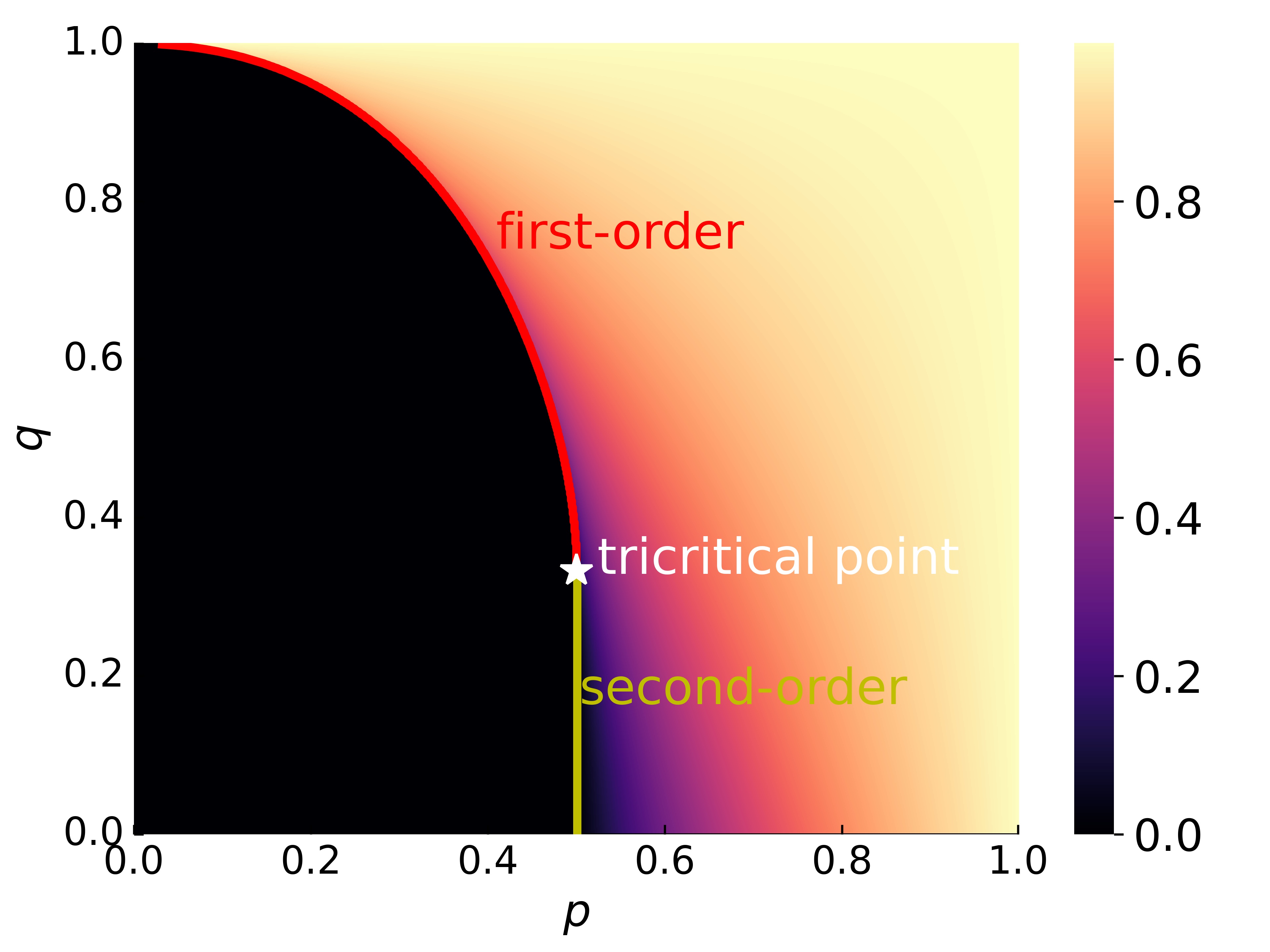} &
    \includegraphics[width=0.32\textwidth]{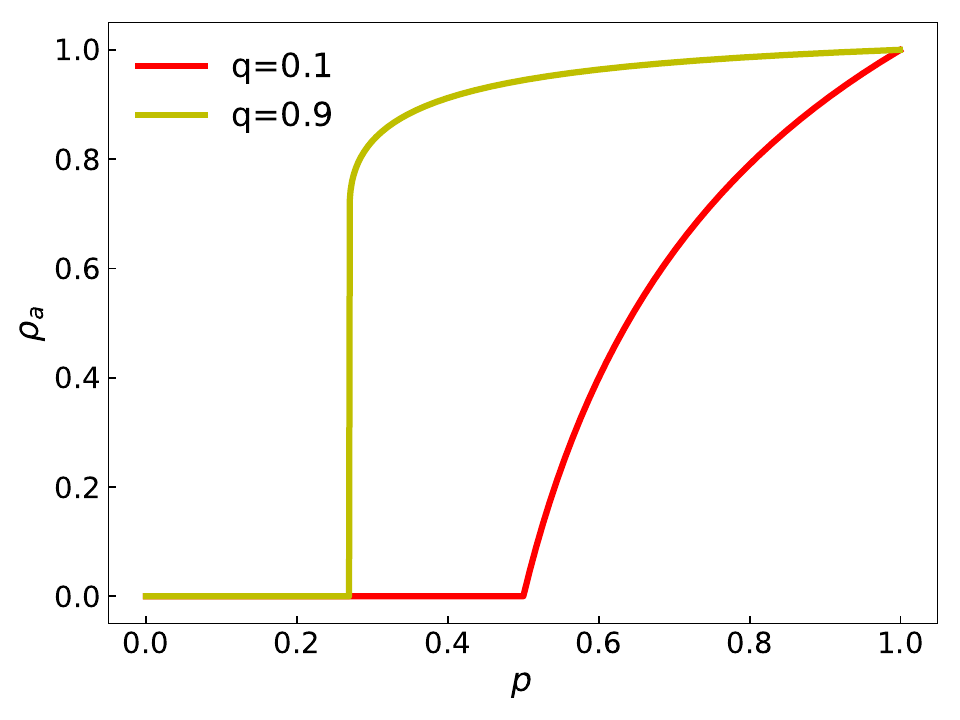} &
    \includegraphics[width=0.32\textwidth]{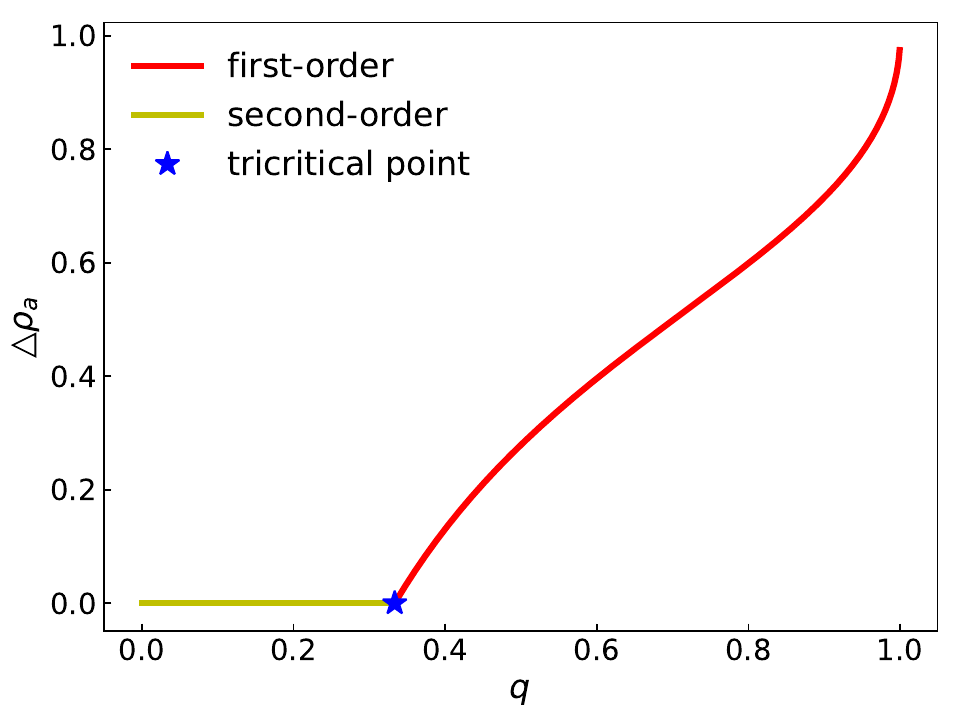} \\
    (a) &  (b)&  (c)\\
\end{tabular}
\caption{Mean-field analysis results for the two-dimensional tricritical directed percolation model. \textbf{(a)} Phase diagram of the TCP model in the mean-field limit, with the tricritical point (white dot) located at $(1/2, 1/3)$. The red line indicates the first-order phase transition, while the green line represents the second-order phase transition, \textbf{(b)} Variation of particle density $\rho_a$ as a function of $p$ for $q = 0.1$ (continuous transition) and $q = 0.9$ (discontinuous transition), calculated using equation~\ref{equ3}, \textbf{(c)} Change in $\Delta \rho_a$ as a function of $q$, computed from equation~\ref{equ5}, highlighting the lines of first-order and second-order phase transitions, as well as the tricritical point.}
\label{meanfiled}
\end{figure*}

To describe the reaction-diffusion process of particles, S. Lübeck proposed the tricritical directed percolation (TDP) model~\cite{lubeck2006tricritical,grassberger2006tricritical}. This model builds on the directed percolation framework by introducing higher-order reaction processes, controlled by a probability parameter $q$. In the TDP model, a particle-pair reaction occurs with probability $q$, generating an active particle in the vicinity of the pair, while with probability $1-q$, the original contact process takes place, where a particle is generated with probability $p$, or transitions from the active to the inactive state with probability $1-p$. The TDP exhibits either first-order or second-order phase transitions depending on the values of the parameters $(p,q)$, where the order parameter undergoes either continuous or discontinuous changes as $p$ varies. A tricritical point $q_{t}$separates these two types of transitions at the boundary between them. Unlike the directed percolation class, the tricritical point in the TDP model belongs to the TDP class.

S. Lübeck identified the critical points and determined the corresponding critical exponents by applying scaling relations of the form $\rho_a \sim \lambda^{-\beta_t} \tilde{r}_q$ to steady-state configurations. However, this approach requires prior knowledge of the tricritical point's location. In contrast, Minjae Jo~\cite{jo2020tricritical} utilized simulations on infinitely large systems to compute the scaling relation $\rho_a(t) \sim t^{-\delta}$, which allowed for the determination of the tricritical point. Jo's work also investigated critical points in long-range correlated models. Nevertheless, Monte Carlo simulations on infinite systems are computationally expensive and still rely on an approximate location of the critical point.They treat the crossover region between first-order and second-order phase transitions as a second-order transition, which only allows them to identify the boundary of the first-order phase transition. Our proposed method can simultaneously distinguish these three regions and determine both the boundary of the first-order phase transition and the crossover region.

In recent years, machine learning methods have been increasingly applied to the study of phase transitions~\cite{arai2018deep,chernodub2020topological,jo2021absorbing,carrasquilla2017machine,hu2017discovering,canabarro2019unveiling,saitta2011phase,dong2019machine,zhang2019machine}. These methods primarily include supervised learning~\cite{canabarro2019unveiling,dong2019machine,zhang2019machine}, unsupervised learning~\cite{hu2017discovering,zhang2019machine,kaming2021unsupervised}, and semi-supervised learning techniques~\cite{hu2017discovering,chen2023study}. Supervised learning, which utilizes labeled datasets, can be trained to identify and predict critical points in phase transitions with high accuracy. For instance, Minjae Jo ~\cite{jo2021absorbing} employed convolutional neural networks (CNNs) to identify critical points in one- and two-dimensional quantum contact processes. Similarly, Shunta Arai ~\cite{arai2018deep} designed a temperature recognition method based on CNNs, where the output was modified to represent temperature. By analyzing the weight changes between the hidden and output layers, they successfully identified the phase boundaries of quantum phase transitions. Unsupervised learning methods, such as principal component analysis (PCA)~\cite{shen2022transfer,zhang2022machine,shen2022machine} and autoencoders~\cite{tuo2024supervised,wang2024supervised,shen2021supervised}, have also been widely used. For example, Zhang~\cite{zhang2022machine} applied PCA and autoencoders to capture critical points and phase transitions in the XY model. Semi-supervised learning methods, which use a partially labeled dataset to identify phase transitions, are also commonly used, with domain-adversarial neural networks~\cite{chen2023study} and Siamese neural networks~\cite{chicco2021siamese, ranasinghe2019semantic, zhang2016siamese} being notable examples. Despite these advancements, no existing method has directly classified types of phase transitions. In this context, this paper proposes a convolutional neural network-based approach to classify phase transition types, while simultaneously identifying the location of the tricritical point.

The main structure of this paper is as follows. In Sec.\uppercase\expandafter{\romannumeral2}, we introduce the reaction processes of the TDP model and its mean-field solution. In Sec.\uppercase\expandafter{\romannumeral3}, Monte Carlo simulations are employed to obtain configurations and the phase diagram, revealing the tricritical point and the extent of the crossover region. Sec.\uppercase\expandafter{\romannumeral4} presents the convolutional neural network method used in this study and discusses the results. Finally, Sec.\uppercase\expandafter{\romannumeral5} concludes the paper with a summary of the findings.

\section{Tricritical directed percolation}

S. Lübeck introduced the TDP model by incorporating higher-order reaction processes into the contact process. The original contact process exhibits nonequilibrium phase transitions, belonging to the directed percolation universality class. With the introduction of higher-order reactions, the model displays both equilibrium and nonequilibrium phase transition behaviors, which are separated by a tricritical point. At the tricritical point, the model exhibits a tricritical universality class distinct from the DP universality class.

The tricritical contact process (TCP) is introduced as a lattice model, similar to the contact process, and takes place on a continuous Markov in a d-dimensional lattice. Each lattice site can either be empty ($n = 0$) or occupied by a particle ($n = 1$). The dynamics are characterized by spontaneous processes occurring at certain transition rates. The CP is described by the probability $p$, where a randomly chosen occupied site generates a new particle in a neighboring site with probability $p$, or undergoes annihilation with probability $1 - p$.

To introduce higher-order reactions, the reaction process is described by the probability $q$. If a randomly chosen occupied site $i$ has an occupied neighbor, a third particle is generated at a randomly chosen empty neighbor of this pair. If site $i$ is isolated, the regular CP update process is executed. When updating a given occupied site $i$, the reaction process is executed with probability $q$; otherwise, the usual CP update steps are followed. The details of the reaction processes are outlined in Table~\ref{Reaction}.

\begin{table}[htbp]
    \centering
    \setlength{\tabcolsep}{2.5pt}
    \begin{tabular}{ccc}
        \hline \hline
              Reaction process       & Probability       & Rate of particle generation             
             \\ \hline
            $A \rightarrow 0$ & $(1-p)(1-q)$ & $-(1-p)(1-q)\rho_a$  \\
            $A0\rightarrow AA$ & $p(1-q)$ & $p(1-q)\rho_a(1-\rho_a)$  \\
            $A0\rightarrow 00$ & $(1-p)q$ & $-(1-p)q\rho_a(1-\rho_a)$   \\
            $A0\rightarrow AA$ & $pq$ & $pq\rho_a(1-\rho_a)$  \\
            $AA0\rightarrow AAA$ & $q$ & $q\rho_a^2(1-\rho_a)$\\ 
            \hline \hline
        \end{tabular}
    \caption{The tricritical contact process model's reaction process and probability.}
    \label{Reaction}
\end{table}

In recent years, non-equilibrium phase transitions have garnered significant attention from researchers in statistical physics. In mean-field analysis, the directed percolation process can be represented by a Langevin equation~\cite{janssen1981nonequilibrium,jo2019nonequilibrium}, where the density of active sites is denoted by $\rho_a$. In the mean-field limit, neglecting the effects of local density fluctuations, the dynamical equation for the TDP model can be written as:
\begin{equation}
    \partial{t}\rho_{a} = (2p-1)\rho_{a}-(p+pq-2q)\rho_{a}^2 - q\rho_{a}^3.
    \label{equ1}
\end{equation}

When the particle density is in a steady state, setting $\partial_t \rho_a = 0$, the above equation (excluding negative solutions) yields the following solution:
\begin{equation}
    \rho_{a} = 0 
    \label{equ2}
\end{equation}
\begin{equation}
    \rho_{a} =\frac{(p+pq-2q)-\sqrt{(p+pq-2q)^2+ 4q(2p-1)}}{-2q}.
    \label{equ3}
\end{equation}
Through linear stability analysis, the steady-state solution $\rho_a = 0$ is stable when $2p - 1 < 0$. For the other steady-state solution, when $\left( p + pq - 2q \right)^2 + 4q(2p - 1) = 0$, the steady-state solution $\rho_a$ has a minimum value given by:
\begin{equation}
    \rho_{a} =\frac{q(2-p)-p}{2q}.
    \label{equ4}
\end{equation}

With a fixed probability $q$, as the probability $p$ increases, the change in particle density, denoted as $\Delta \rho_a$, describes the transition from $\rho_a = 0$ to $\rho_a > 0$. When $q < 1/3$, the change $\Delta \rho_a = 0$, indicating a second-order phase transition. In contrast, when $q > 1/3$, $\Delta \rho_a > 0$, signaling a first-order phase transition. Therefore, $q = 1/3$ marks the tricritical point of the TCP model. The expression for $\Delta \rho_a$ is given by:
\begin{equation}
    \Delta\rho_{a} =
    \begin{cases}
    \centering
    \qquad\qquad 0&q\le\frac{1}{3}\\
    1-\frac{\sqrt{\frac{3q+1}{q}(1-q)}+q-1}{q+1} &q>\frac{1}{3}.
    \end{cases}
    \label{equ5}
\end{equation}

Using equation~\ref{equ3}, we calculated the particle density of the system for various values of $(p, q)$, resulting in the phase diagram shown in Fig~\ref{meanfiled}(a). The diagram reveals both first-order and second-order phase transition boundaries, along with the tricritical point where these boundaries intersect. For $q < 1/3$ (as shown in Fig~\ref{meanfiled}(b) for $q = 0.1$), the model exhibits a second-order phase transition, with particle density increasing from zero as $p$ rises. In contrast, for $q > 1/3$ (illustrated in Fig~\ref{meanfiled}(b) for $q = 0.9$), the model demonstrates first-order phase transition behavior, characterized by a sudden jump in particle density from a value greater than zero as $p$ increases. Fig~\ref{meanfiled}(c) shows the variation of $\Delta \rho_a$ as a function of $q$, calculated using equation~\ref{equ5}. This allows for the straightforward identification of the first-order and second-order phase transition lines, as well as the position of the tricritical point. Notably, $\Delta \rho_a$ increases approximately linearly with $q$.

Next, we performed Monte Carlo simulations to describe the TCP model. These simulations were carried out on a fully connected two-dimensional lattice to numerically verify the phase diagram. By calculating the particle density at active sites from the simulation results, we confirmed the existence of the tricritical point. Furthermore, we identified the location of the tricritical point through the particle density increments, while also providing configurations for use in machine learning methods.

\section{Monte Carlo simulations}

We performed numerical simulations based on the dynamics of the TCP model described above. Specifically, the model is implemented on a two-dimensional lattice, where each site can be either in an active state (denoted as 1) or an inactive state (denoted as 0). In this study, the initial configuration consists of half the sites being active and the other half being inactive. Periodic boundary conditions are applied throughout the simulation.

A single time step is defined as $N = L^2$, where $L$ is the lattice size, and during each time step, the following rules are applied. A random active site $i$ is selected, and with probability $q$, a two-particle reaction process occurs: a neighboring site $j$ is randomly chosen, and if it is occupied, one of the six neighbors of the pair is selected to become active. If the chosen neighbor $j$ is inactive, it becomes active with probability $p$.With probability $1 - q$, the regular CP process occurs: a random active site $i$ is selected, and a random neighbor site $j$ is chosen. If $j$ is occupied, it becomes inactive with probability $1 - p$; if site $j$ is unoccupied, it becomes active with probability $p$.

\begin{figure}[!thb]
\centering
    \includegraphics[width=0.48\textwidth]{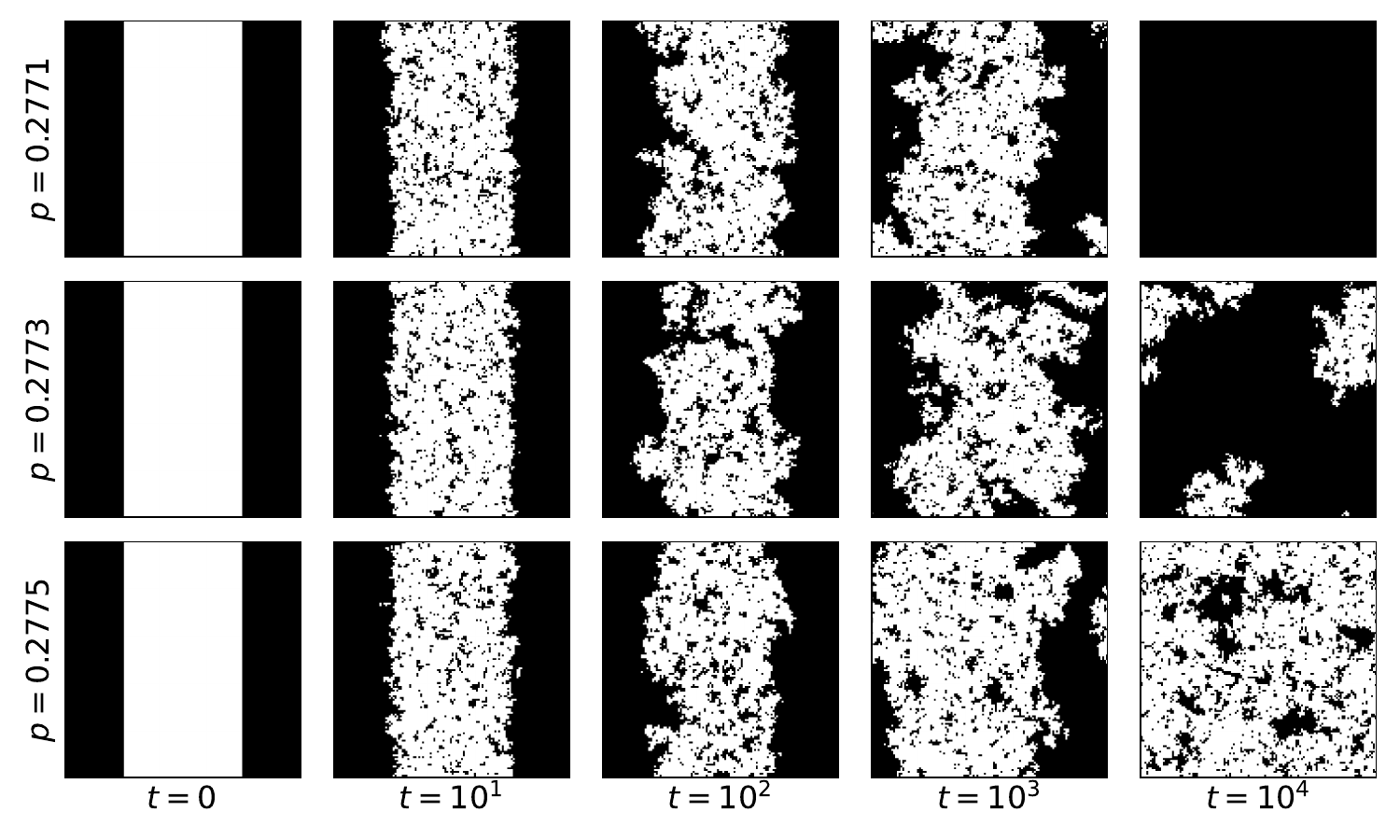} 
\caption{Configurations at different time steps for a system of size $L = 128$. \text All configurations are taken with $q = 0.97$. The values of $p$ from top to bottom are $p = 0.2771$, $p = 0.2773$, and $p = 0.2775$. The time steps from left to right are $t = 0$, $t = 10$, $t = 100$, $t = 1000$, and $t = 10000$.}
\label{configuration}
\end{figure}

Based on the simulation process, we obtained various configurations. Fig~\ref{configuration} displays the configurations of the two-dimensional TCP model with a lattice size of $L = 128$ at $q = 0.97$ for five different time steps. Near the critical point, achieving a steady-state configuration requires an increasing number of time steps, which is further amplified by the system size. Consequently, large-scale models consume significantly more computational resources.

\begin{figure*}[htbp]
\centering
\begin{tabular}{ccccccccc}   
    \includegraphics[width=0.32\textwidth]{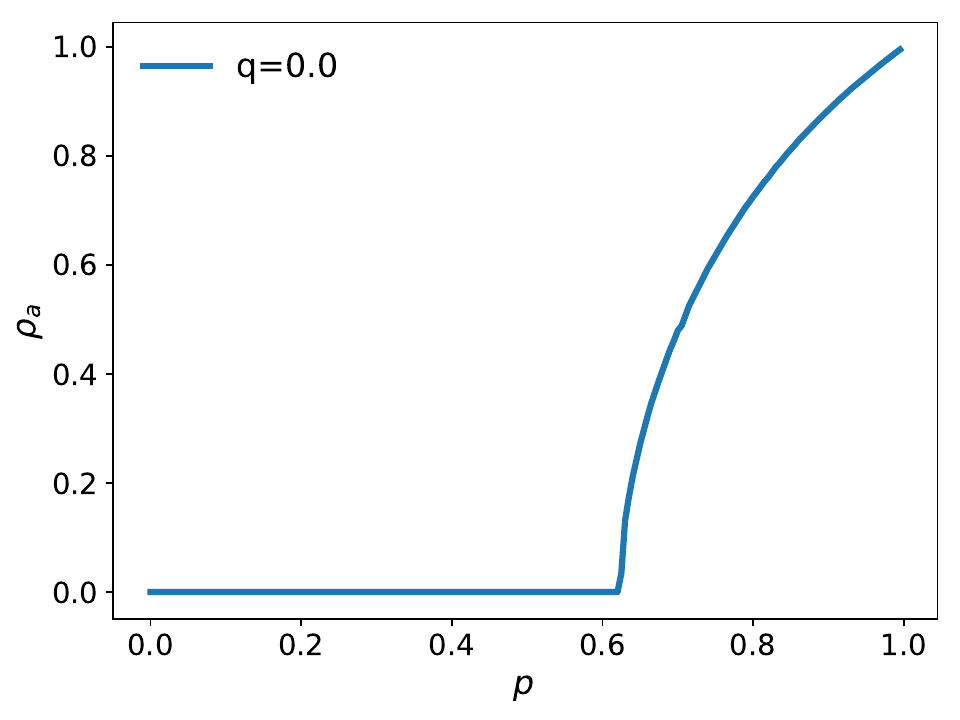} &
    \includegraphics[width=0.32\textwidth]{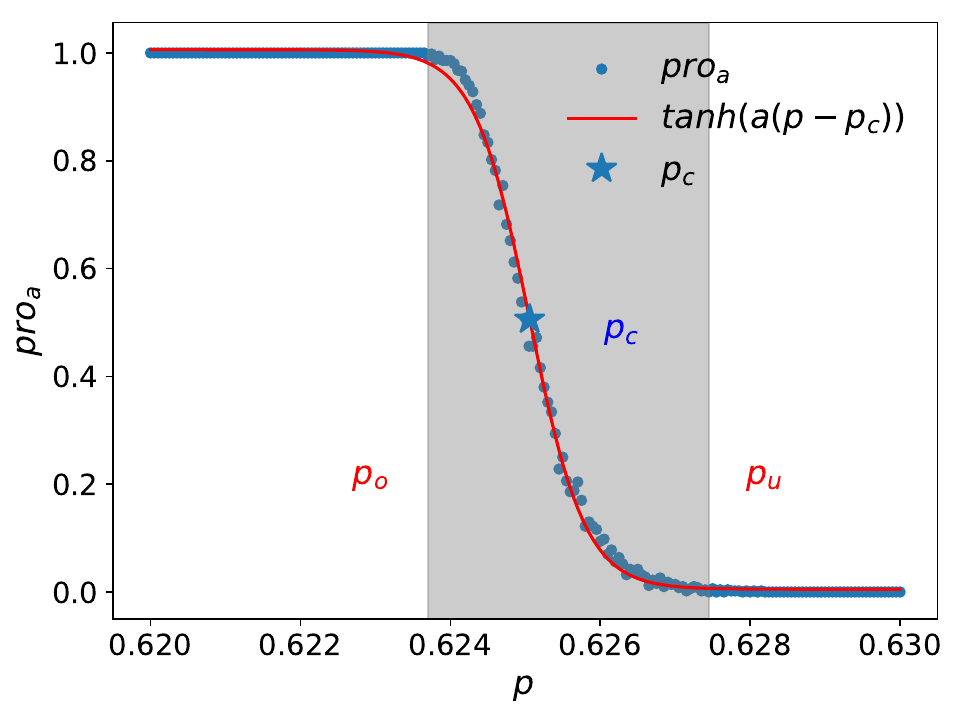} &
    \includegraphics[width=0.32\textwidth]{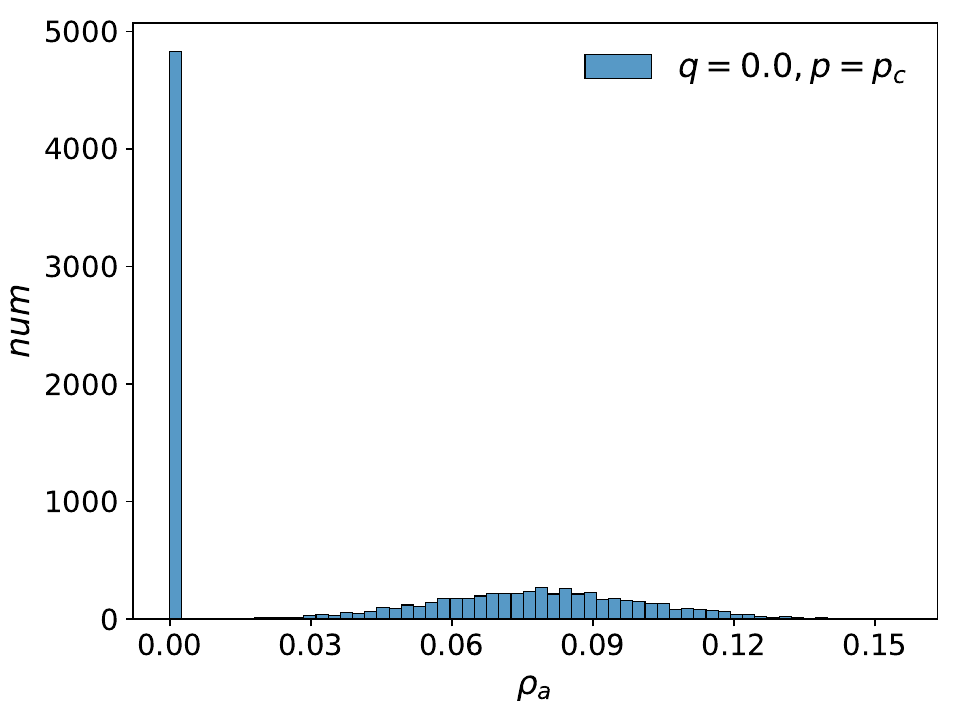} &\\
    (a) &  (b)&  (c)\\
    \includegraphics[width=0.32\textwidth]{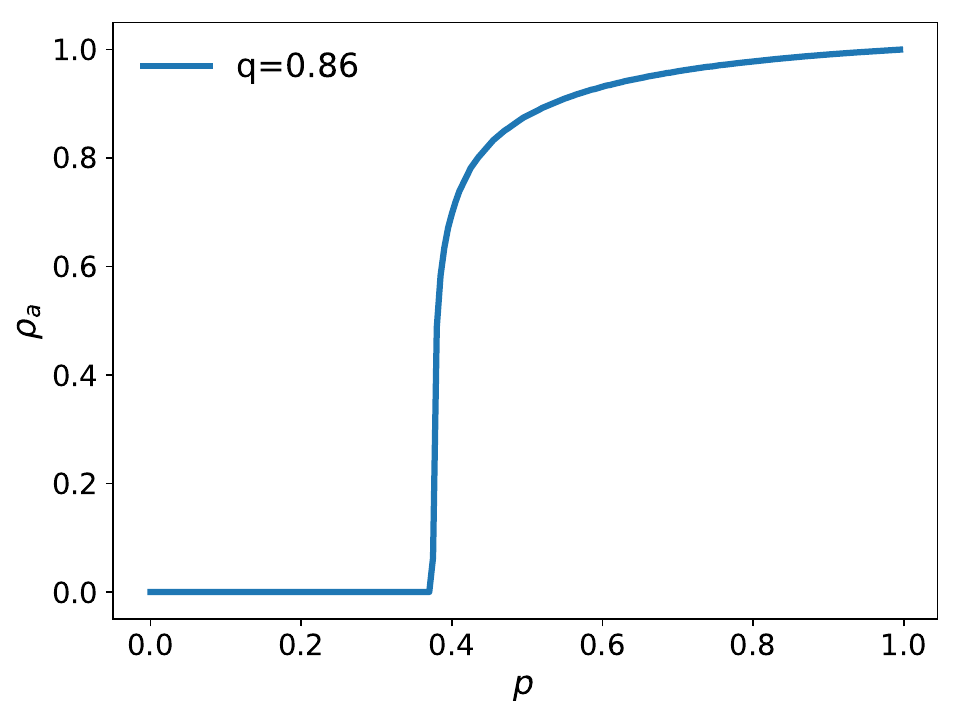} &
    \includegraphics[width=0.32\textwidth]{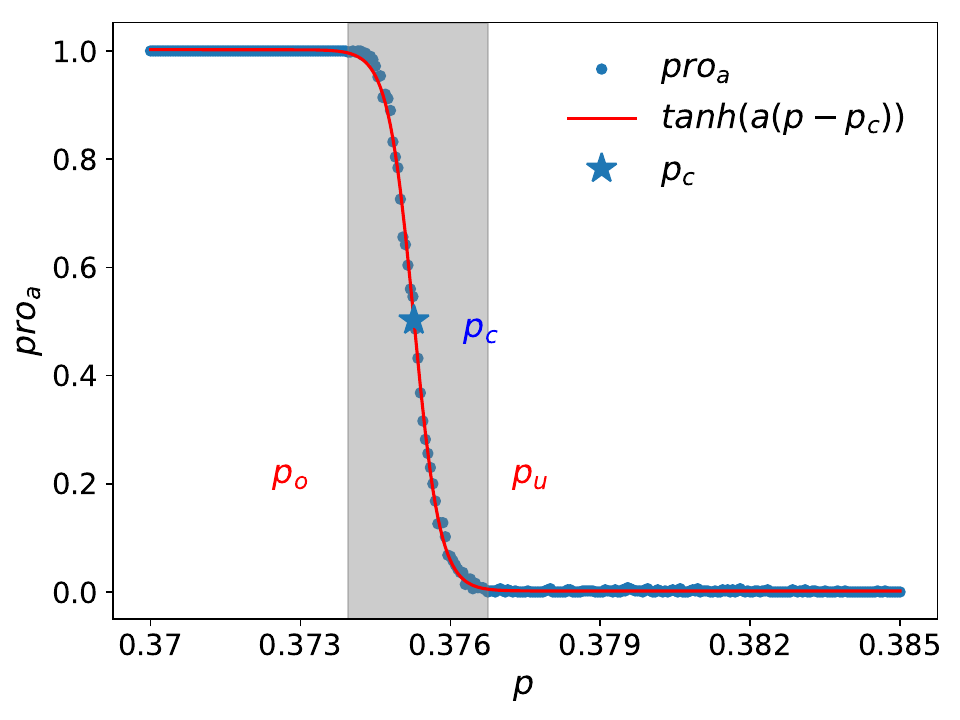} &
    \includegraphics[width=0.32\textwidth]{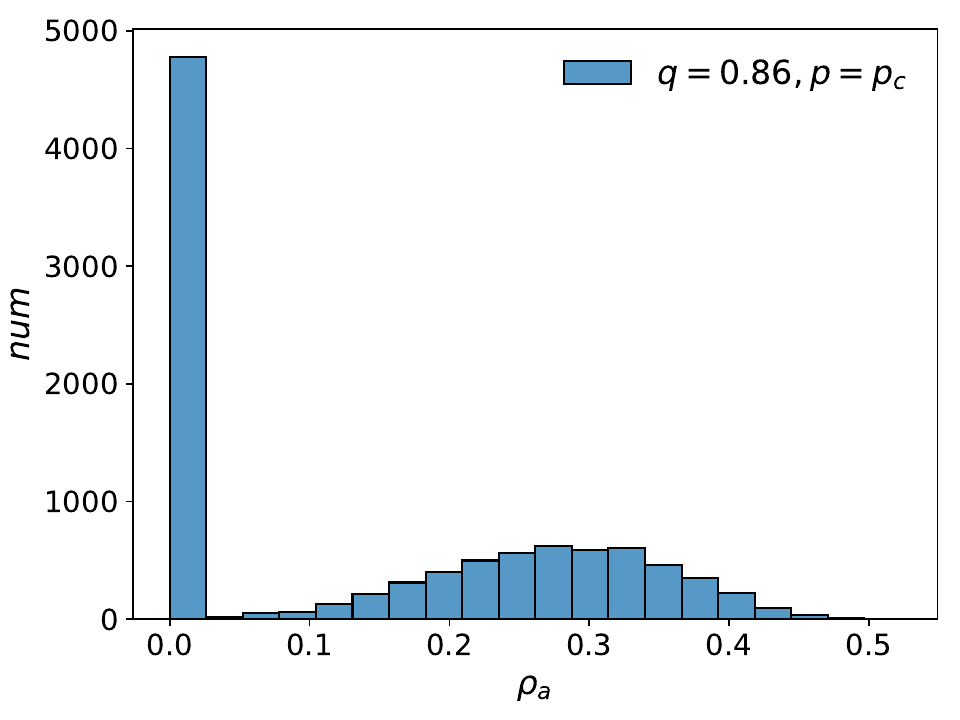} &\\
     (d)&  (e)&  (f)\\
    \includegraphics[width=0.32\textwidth]{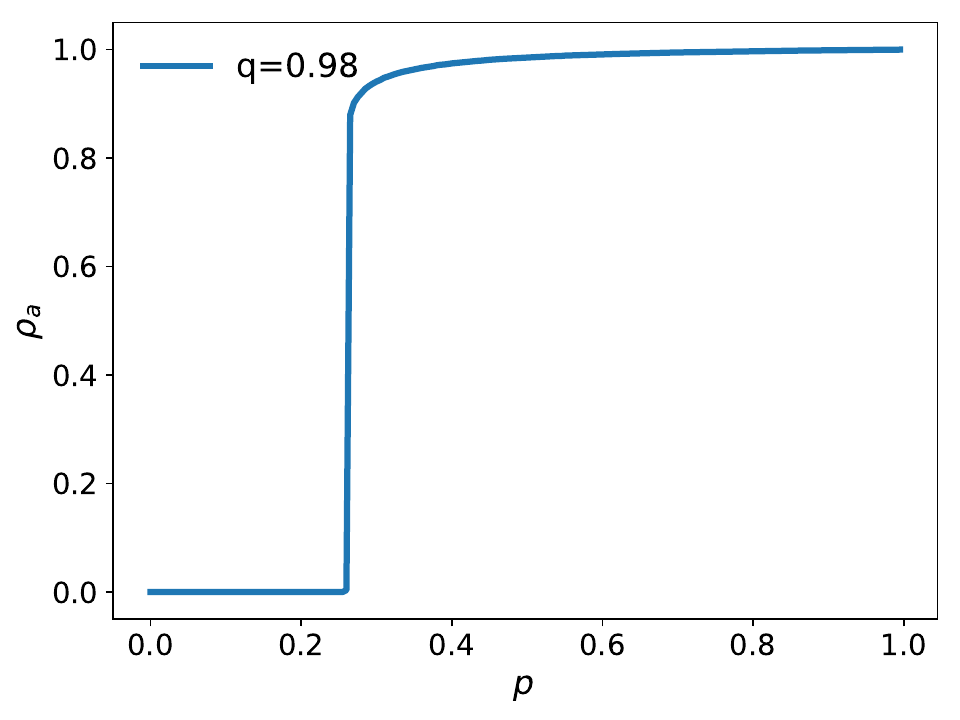} &
    \includegraphics[width=0.32\textwidth]{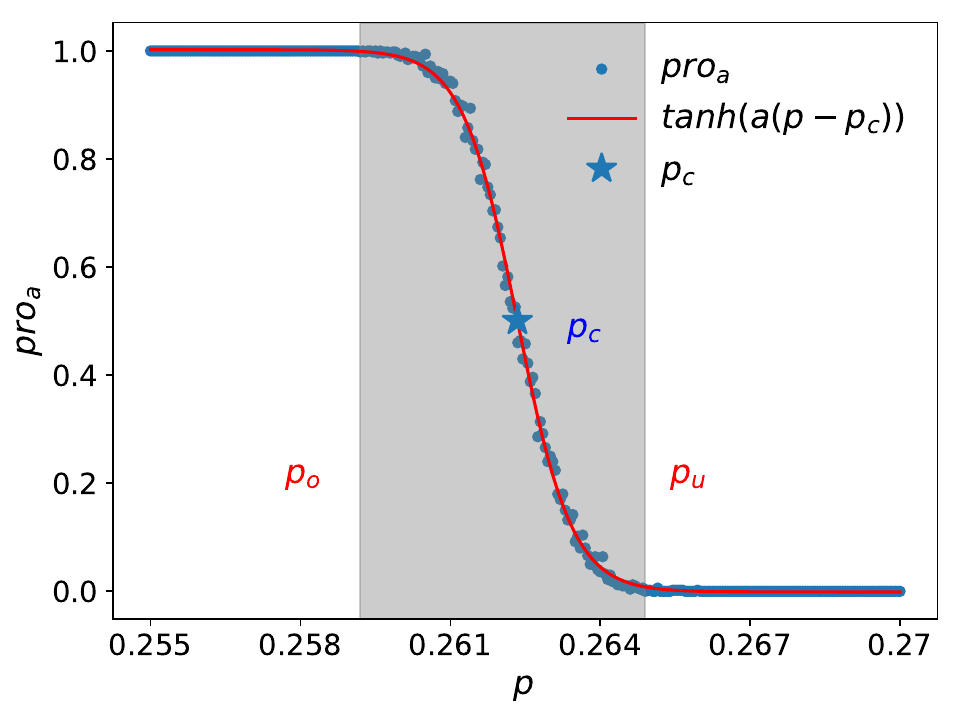} &
    \includegraphics[width=0.32\textwidth]{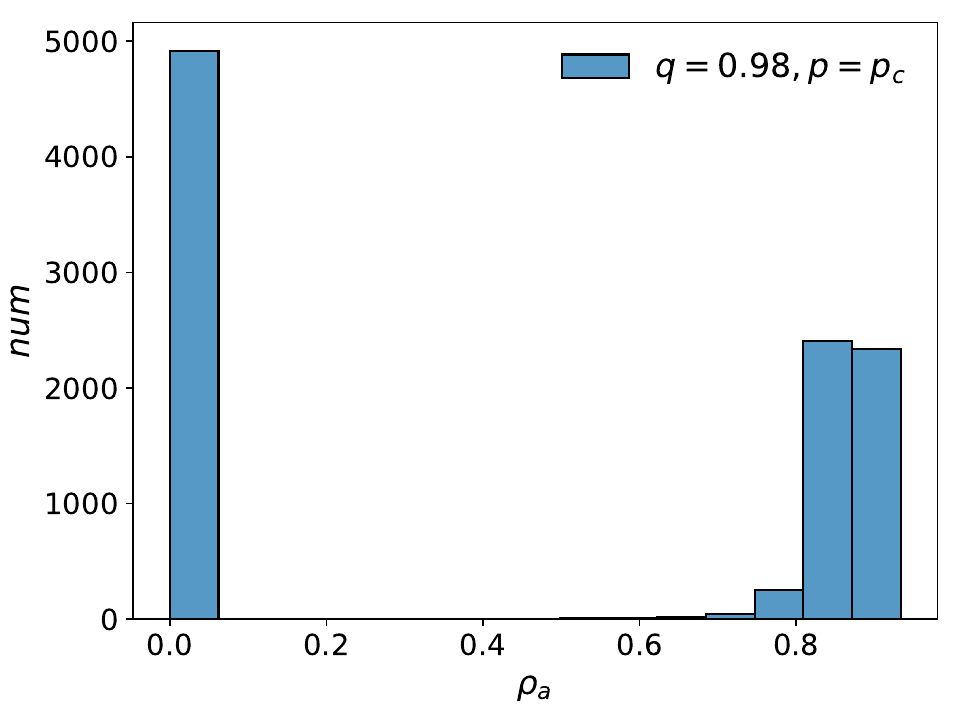} & \\
     (h)&  (i)&  (j)\\
\end{tabular}
\caption{Behavior of particle density under different $q$ values. From top to bottom, the values of $q$ are $q=0.0$, $q=0.86$, and $q=0.98$.Panels \textbf{(a)}, \textbf{(d)}, and \textbf{(h)} illustrate the changes in particle density within the jump interval as $p$ varies.Panels \textbf{(b)}, \textbf{(e)}, and \textbf{(i)} show the probability of the absorbing phase as a function of $p$.with curve fitting used to determine the critical point $p_c$ and the mixed interval $[p_o,p_u]$ of absorbing and non-absorbing phases(Grey area). For example, in panel \textbf{(b)}, we find $p_c=0.62506$ for $q=0.0$, with the mixed interval given as $[0.6237,0.62745]$.Panels \textbf{(c)}, \textbf{(f)}, and \textbf{(j)} present the statistical results of the particle density from the Monte Carlo simulations.}
\label{mc}
\end{figure*}

For instance, when $p = 0.2775$, the system reaches a steady state characterized by a relatively high particle density. In contrast, at $p = 0.2771$, the system enters an absorbing state with $\rho_a = 0$, indicating a first-order phase transition at high $q$. This illustrates that, near the critical point, the value of $p$ significantly influences the system's steady-state density.

We examined the system by sampling 10,000 points in the parameter space of $q = 0:0.02:1$ and $p = 0:0.02:1$. For each pair of $(p, q)$, the simulations were repeated 500 times to calculate the average steady-state density. For larger values of $q$ (as shown in Fig~\ref{mc}(a) for $q = 0.0$), the particle density $\rho_a$ exhibits a discontinuous jump at a certain value of $p$, characteristic of a first-order phase transition. Conversely, for smaller values of $q$ (illustrated in Fig~\ref{mc}(h) for $q = 0.98$), the phase transition is continuous, indicating a second-order phase transition. Notably, when $q$ is close to the tricritical point $q_t$ (as in Fig~\ref{mc}(d) for $q = 0.86$), distinguishing the phase transition type becomes more challenging based on particle density changes.

In the previous step, we varied $q$ from 0 to 1 in increments of $\Delta q = 0.02$ and $p$ from 0 to 1 in increments of $\Delta p = 0.02$, performing 500 Monte Carlo simulations to calculate the average steady-state particle density. However, this approach did not allow us to distinguish the phase transition types near the critical point. Therefore, we proceeded to identify the jump intervals for particle density at each value of $q$ (from $\rho_a = 0$ to $\rho_a > 0$). For instance, at $q = 0.0$, the jump interval is [0.62, 0.63]. Within these intervals, we defined $\Delta p = 0.0005$ and conducted further fine-grained simulations of 500 runs to compute the particle density and the probability of the absorbing state. The third step involved locating the mixed interval $[p_o, p_u]$ where both absorbing and non-absorbing phases coexist, as illustrated in Fig~\ref{mc}(b),Fig~\ref{mc}(e) and Fig~\ref{mc}(i).

For any value of $q$, there exists a region $[p_o, p_u]$ where the system exhibits a mixture of absorbing and non-absorbing states in its steady state. Within this region, we calculated the probability of the absorbing phase, denoted as $pro_a$, using Monte Carlo simulations.By fitting the data with a hyperbolic tangent function $tanh(a(p-p_c))+b$, we determined the critical value $p_c$ which is a function of $q$ and is denoted as $p_c(q)$. As shown in Fig~\ref{mc}(b), the curve fitting yields $p_c=0.62506$ for $q=0.0$, consistent with the theoretical solution of the directed percolation model~\cite{hinrichsen2000non}.

For each different $q$, we performed 10,000 Monte Carlo simulations at the critical state $(p=p_c)$to obtain the steady-state density distribution. In Fig~\ref{mc}(j) for $q=0.98$ , we observe a portion of the absorbing phase alongside a significant non-absorbing phase, indicating a first-order phase transition as the system transitions from the absorbing state to a non-absorbing state with higher particle density. Conversely, in Fig~\ref{mc}(c) for $q=0.0$, the system exhibits an absorbing phase with a continuous non-absorbing phase, characteristic of a second-order phase transition. However, for $q=0.86$ shown in Fig~\ref{mc}(f), it is challenging to accurately determine the type of phase transition, highlighting the importance of identifying the tricritical point.

\begin{figure}[htbp]
\centering
\includegraphics[width=0.45\textwidth]{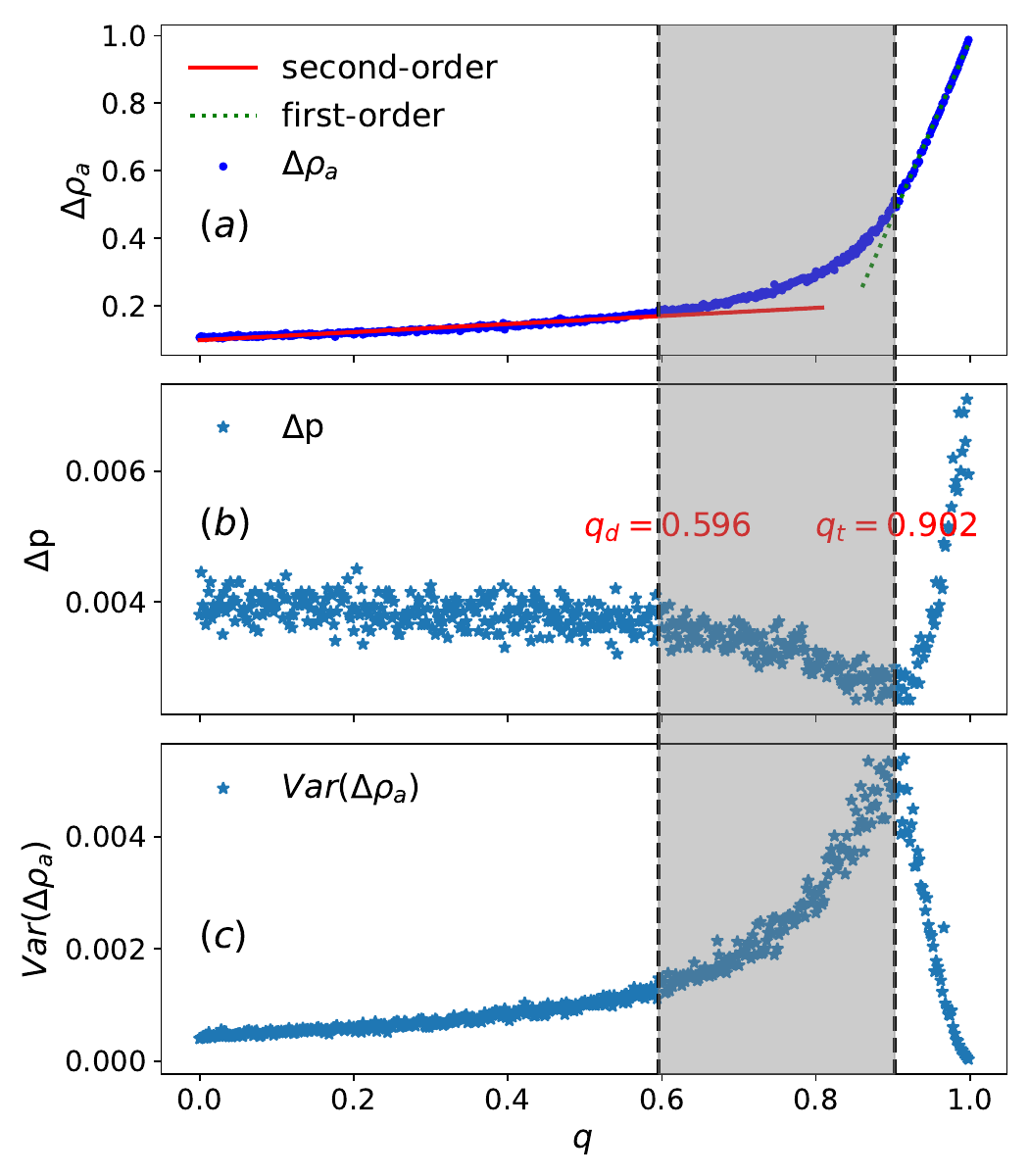}
\caption{Phase transition boundary points. \textbf{(a)} shows the variation of 
$\Delta \rho_a$ for different values of $q$. When $q$ is small, the system exhibits a continuous phase transition; when $q$ is large, it displays first-order phase transition behavior. \textbf{(b)} illustrates the magnitude of $\Delta p=p_u-p_o$. \textbf{(c)} presents the size of $Var(\Delta \rho_a)$. From the figure, distinct regions are evident: a first-order phase transition region($q>q_t$), a second-order phase transition region($q<qd$), and a crossover region between the two types of phase transitions (shaded gray area). At the tricritical point $q=q_t$, $\Delta \rho_a$ is minimized, while $Var(\Delta \rho_a)$ reaches its maximum.}
\label{gap}
\end{figure}

To accurately determine the position of the tricritical point, we define the quantity $\Delta \rho_a = \rho_a(p = p_u) - \rho_a(p = p_o)$ as the difference in the average particle density between the fully absorbing state and the fully non-absorbing state. This represents the difference in the steady-state average particle density across the mixed phase region.

As shown in the Fig~\ref{gap}(a), $\Delta \rho_a$ increases with $q$. For small values of $q$, $\Delta \rho_a$ follows a relatively gentle linear relationship with $q$, indicating a second-order phase transition, where the particle density increases smoothly with $p$. In contrast, for larger values of $q$, $\Delta \rho_a$ exhibits a steeper linear relationship, characteristic of a first-order phase transition, where the particle density undergoes a discontinuous jump.

By fitting the data with straight lines, we observe that the slopes differ between the two regimes. The endpoints of these two linearly fitted straight lines form (gray filled area in the Fig~\ref{gap}) a region corresponding to the crossover effect between the two phase transitions. This interaction zone reflects the difficulty in precisely identifying the tricritical point in the TCP model, as it marks the transition from second-order to first-order phase behavior.

We calculated the mean and variance of $\Delta \rho_a$ for different values of $q$, as well as the corresponding values of $\Delta p$. As shown in the Fig~\ref{gap}(b) and Fig~\ref{gap}(c), the first dashed line marks the endpoint of the second-order phase transition at $q_d = 0.596$, while the second dashed line represents the starting point of the first-order phase transition at $q_t = 0.902$, which also corresponds to the tricritical point.

Between the first-order and second-order phase transitions, there exists an crossover region, as illustrated by the size of the mixed-phase region in the figure below. The presence of cross-over effects complicates the precise determination of the critical point. However, the method used in this study allows for an accurate identification of the tricritical point without prior knowledge of its location and also reveals the extent of the crossover region.

\section{neural network}
\subsection{neural network result of $p_c$}

\begin{figure*}[htbp]
\centering
\includegraphics[width=0.8\textwidth,trim=10 220 10 10,clip]{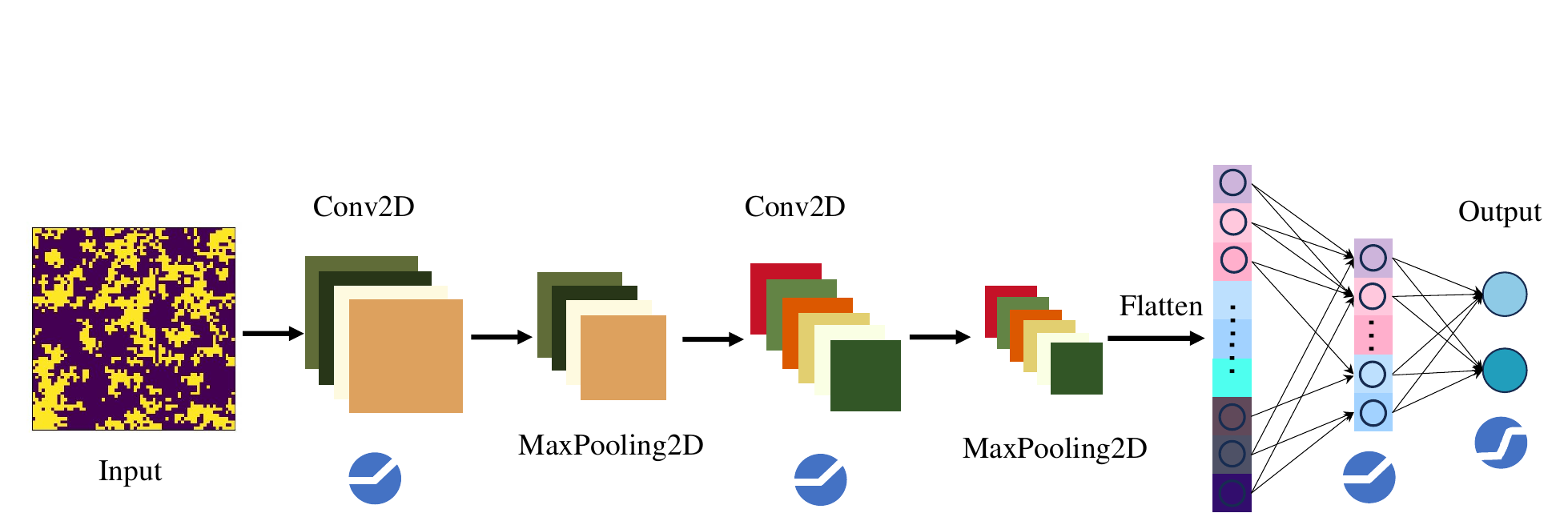}
\caption{Schematic illustration of the convolutional NN.}
\label{cnn}
\end{figure*}

\begin{figure*}[htbp]
\centering
\begin{tabular}{cccccc}   
    \includegraphics[width=0.32\textwidth]{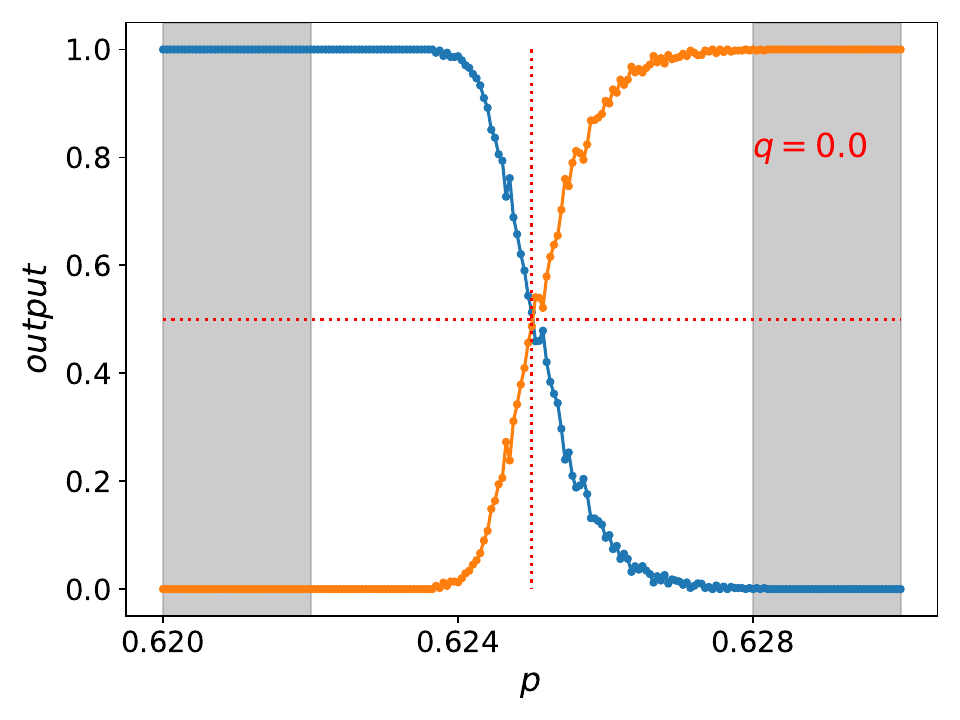} &
    \includegraphics[width=0.32\textwidth]{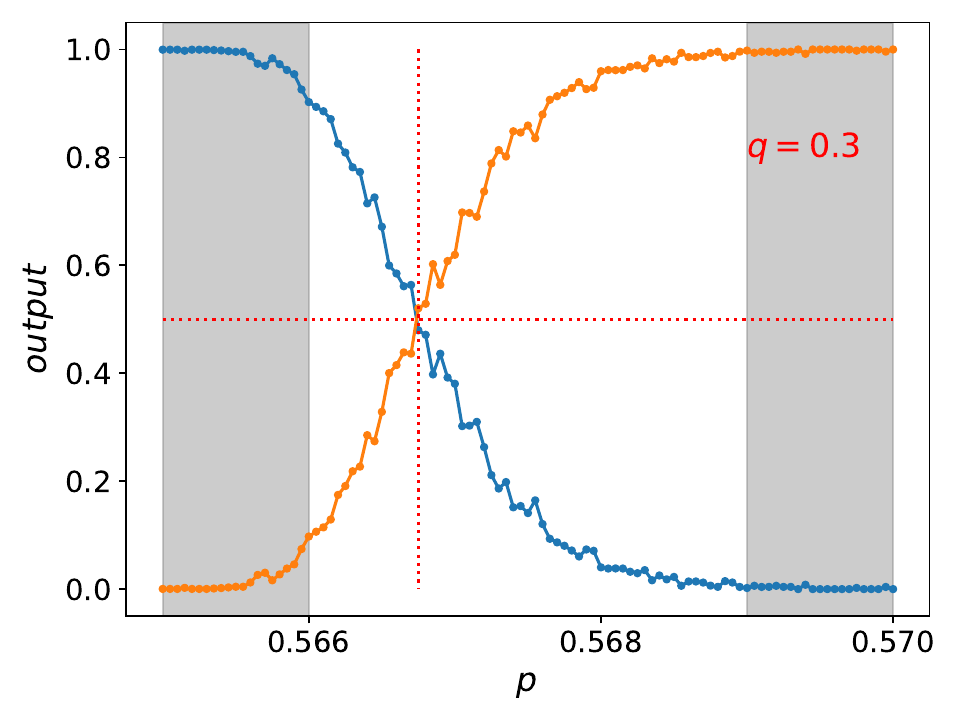} &
    \includegraphics[width=0.32\textwidth]{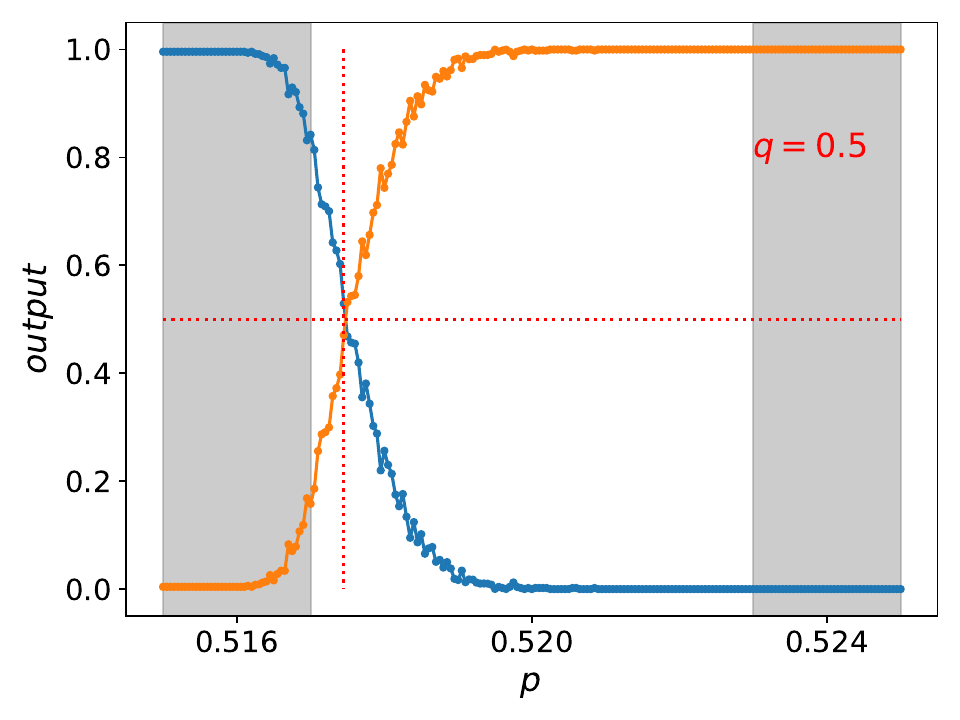} &\\
    (a) &  (b)&  (c)\\
    \includegraphics[width=0.32\textwidth]{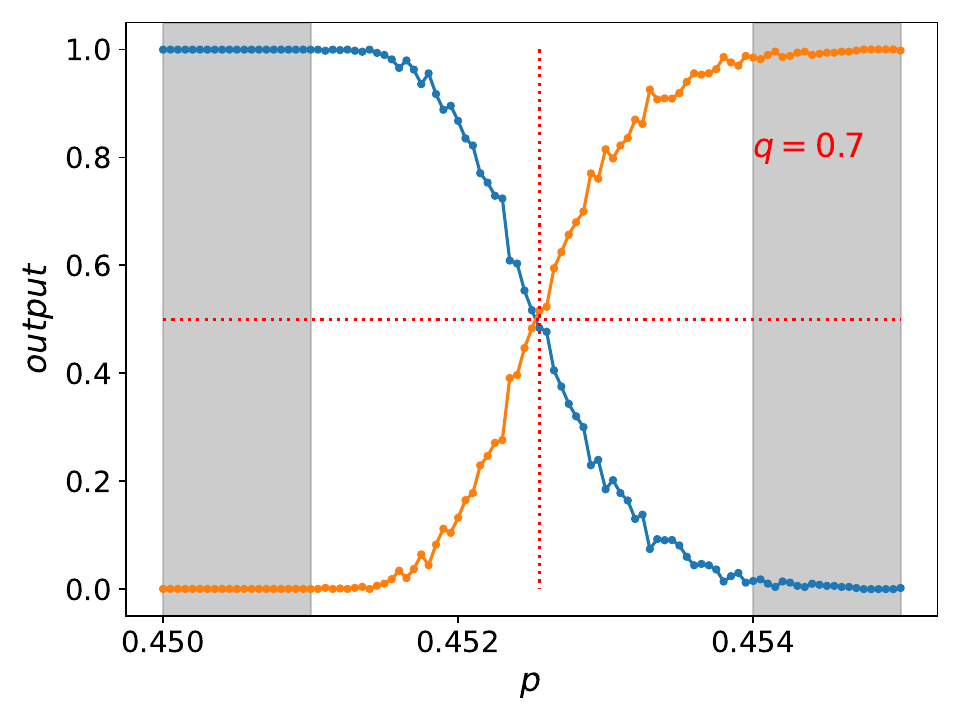} &
    \includegraphics[width=0.32\textwidth]{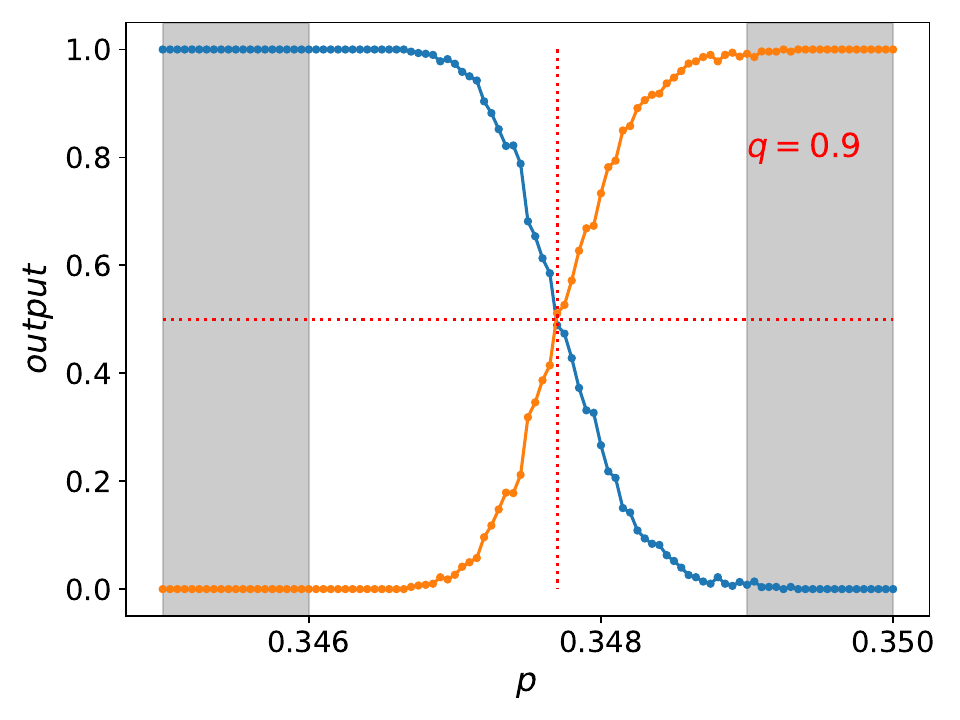} &
    \includegraphics[width=0.32\textwidth]{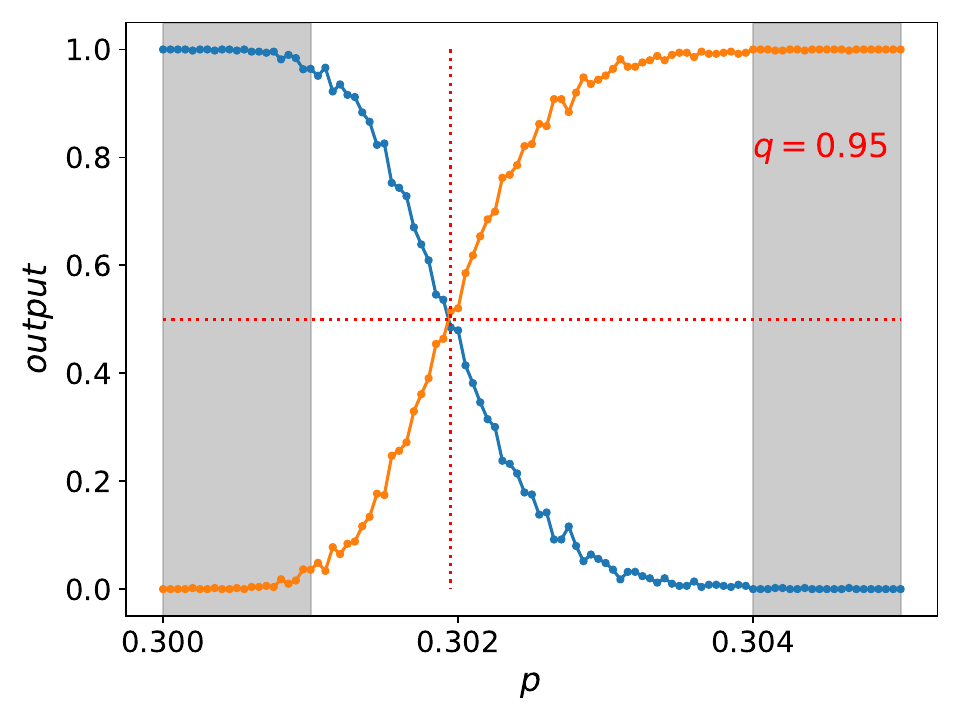} &\\
     (d)&  (e)&  (f)\\
\end{tabular}
\caption{Plots obtained using NN approach.The figures correspond to $q=0.0,0.3,0.5,0.7,0.9$, and $0.95$. For each value of $q$, the average outputs from the neural network on the test set are presented as functions of $p$. The intersection points of the two outputs indicate the critical point $p_c$. The gray shaded regions represent the dataset labeled by the neural network.}
\label{nn_pc}
\end{figure*}

We first used Monte Carlo simulations to generate 500 steady-state configurations on a $64 \times 64$ lattice. For different values of $q$, we performed 5000 iterations, where $q_n = n \Delta q$, with $n = 1, 2, \dots, 500$ and $\Delta q = 0.002$, yielding 500 points in the range from $0$ to $1$. For each specific $q_n$, we also varied $p$ within the interval $p_o:0.00005:p_u
$, corresponding to the mixed interval at a given $q$. For each $(q, p)$ pair, the simulation was run for $t = 5000$ time steps, ensuring the system reached its steady state before storing the resulting configuration. These configurations were then used as input for the neural network.

Using the Keras machine learning library with TensorFlow~\cite{abadi2016tensorflow}, we built a convolutional neural network (CNN), as shown in the Fig~\ref{cnn}. The network consists of an input layer, convolutional layers, and fully connected layers. The configurations generated by Monte Carlo simulations are processed through convolution and pooling operations and then passed as input to a three-layer fully connected network. The three fully connected layers have 256, 64, and 2 nodes, respectively, all using the rectified linear unit ReLU as the activation function. The output layer consists of 2 nodes with a softmax activation function.

\begin{figure}[h]
\centering
\includegraphics[width=0.45\textwidth]{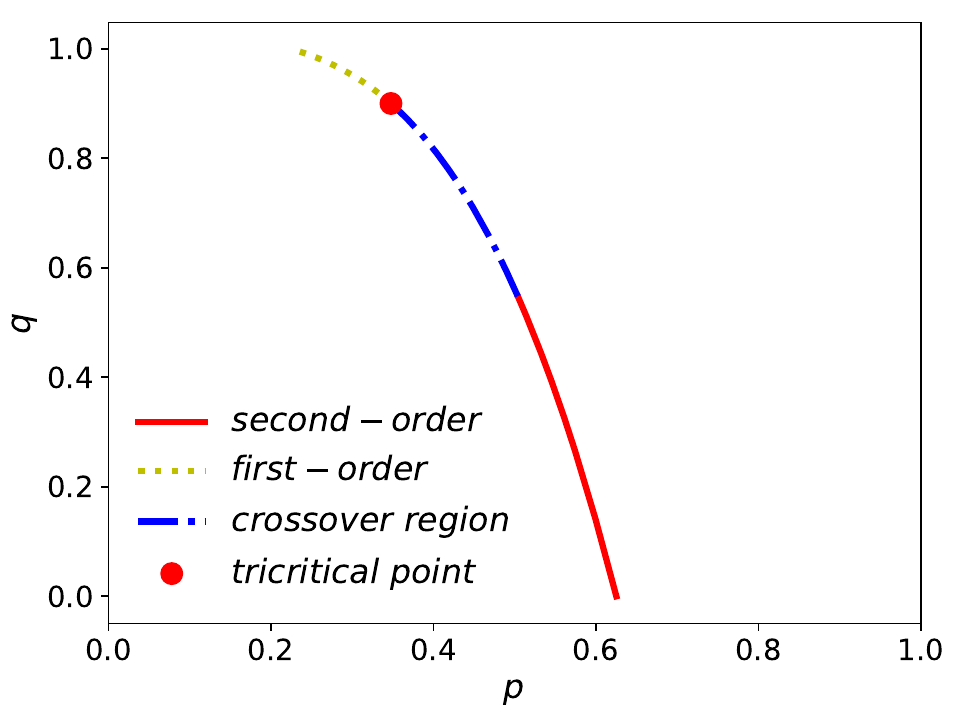}
\caption{The neural network obtains the phase transition lines and the tricritical point diagram.}
\label{pq_ans}
\end{figure}

We set up the training and test sets with $p$ values in the ranges $[0: 0.2p_o]$ and $[0.8p_u: p_u]$, where configurations with $p \in [0: 0.2p_o]$ were labeled as 0 and those with $p \in [0.8p_u: p_u]$ were labeled as 1, as shown in the Fig~\ref{nn_pc}. After training the network, we used the entire range of $p$ values as input and computed the average output of the two neurons for each different $p$. The intersection point of the two output neurons corresponds to the critical point $p_c$. By applying this method for different values of $q$, we determined $p_c$ as a function of $q$, as shown in the Fig~\ref{pq_ans}.

\subsection{neural network result of $q_t$}

Using the convolutional neural network constructed in the previous step, we obtained the function $p_c(q)$. Next, to determine the tricritical point, we constructed another convolutional neural network with the same architecture. However, unlike the previous step, the input configurations for this network were the steady-state configurations at $(q, p_u)$, with the values of $p_u$ determined in the previous step for $q = 0:0.02:1$.For the training and test sets, we set $q \in [0:0.2]$ and $q \in [0.9:1]$, labeling configurations with $q \in [0:0.2]$ as 1 and those with $q \in [0.9:1]$ as 0. After training the network, we calculated the average output for all values of $q$.

\begin{figure}[t]
\centering
\includegraphics[width=0.45\textwidth]{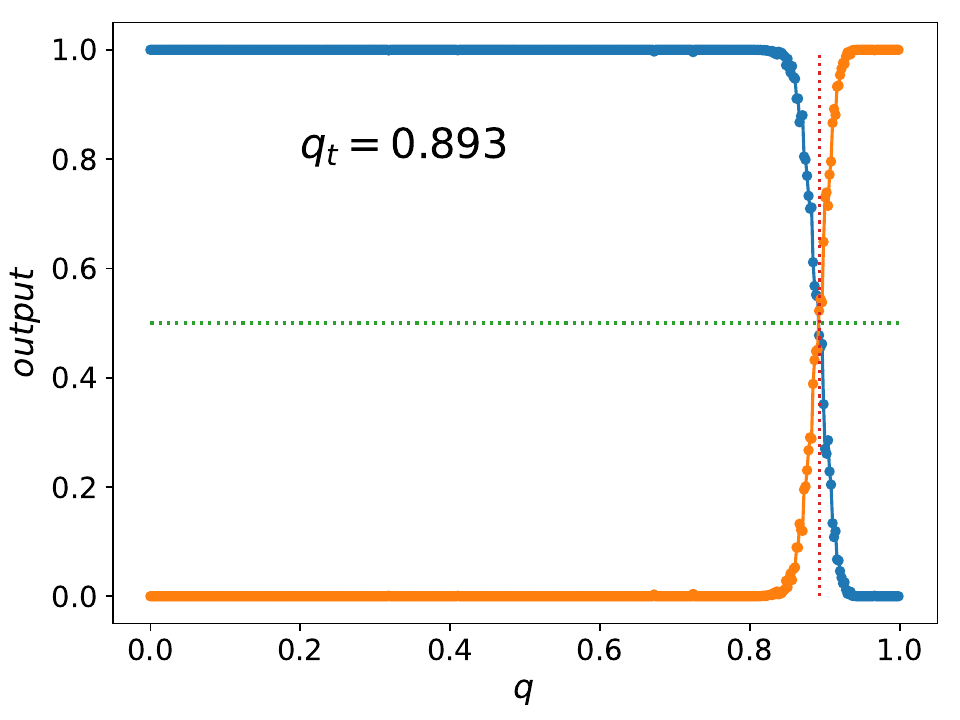}
\caption{The result diagram of the neural network solving for $q_t$.}
\label{qt}
\end{figure}

The intersection of the two output neurons corresponds to the tricritical point, which was found to be $q_t = 0.893$. This result is consistent with the tricritical point identified through Monte Carlo simulations.By using the neural network to learn from the input configurations, we were able to determine the critical points $p_c(q)$ for different values of $q$, and through the analysis of steady-state configurations for $(q, p_u)$, we identified the existence of the tricritical point, as shown in the Fig~\ref{qt}.

\subsection{results of neural network application in determining $q_d$ and $q_t$}
To simultaneously identify the two boundary points shown in Fig~\ref{gap}, we constructed a three-output convolutional neural network. The network architecture is similar to that in Fig~\ref{cnn}, with an additional output. Based on the results from Fig~\ref{gap}, we know that the model exhibits a continuous phase transition for $q \in [0:q_d]$ a first-order phase transition for $q \in [q_t:1]$ and an interaction region between the two types of phase transitions for $q \in [q_d:q_t]$. To differentiate these three regions, we employed a three-output supervised learning approach.

\begin{figure}[h]
\centering
\includegraphics[width=0.45\textwidth]{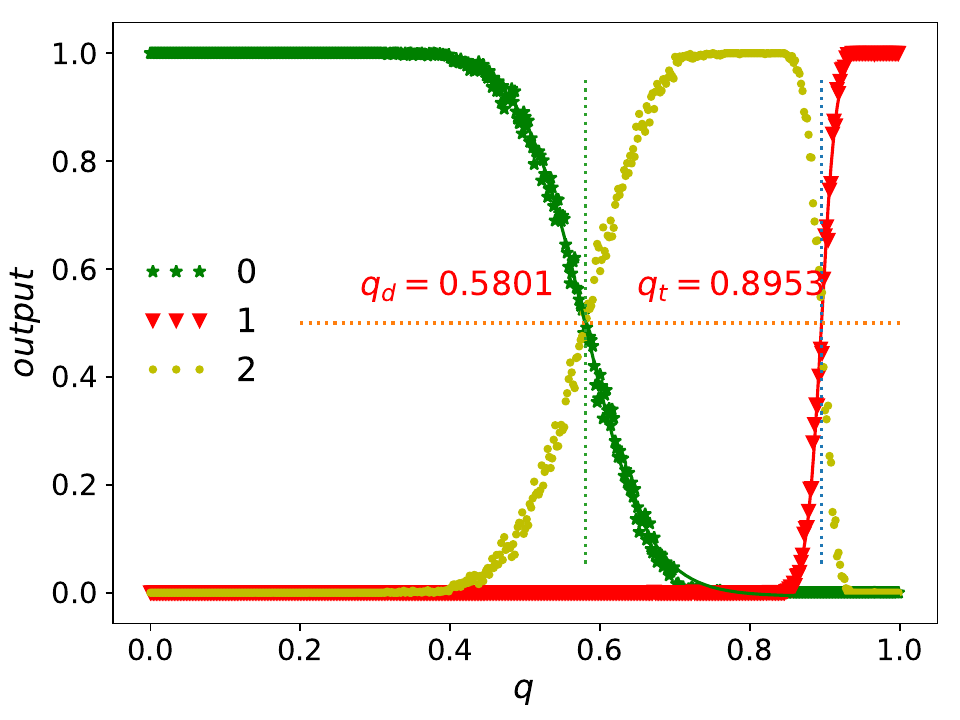}
\caption{Diagram illustrating the neural network's simultaneous identification of $q_d$ and $q_t$.The network is configured with three outputs corresponding to: first-order phase transition, second-order phase transition, and the interaction region between the two. After training, the network identifies the two intersection points as $q_d=0.5801$ and $q_t=0.8953$.}
\label{3phase}
\end{figure}

The training and testing sets were labeled such that $q$ in the range $[0:0.2]$ corresponds to 0, $q$ in the range $[0.92:1]$ corresponds to 1, and $q$  in the range $[0.7:0.8]$ corresponds to 2. Each unique $q$ included 300 samples in the training set and 200 samples in the testing set. After training the neural network, we input all samples to obtain the average output for each $q$. As shown in Fig~\ref{3phase}, the network generates three outputs, resulting in two intersection points at $q_d=0.5801$ and $q_t=0.8953$, which align closely with the results from Monte Carlo simulations. However, since the boundary points from the Monte Carlo method were derived through linear fitting, there may be some deviation in the estimate for $q_d$.

Furthermore, the flexibility of the neural network model allows it to handle large-scale datasets, providing greater computational efficiency and accuracy. This approach offers new tools for future research, particularly in exploring non-equilibrium phase transitions and critical phenomena. Overall, these findings not only enhance our understanding of the tricritical point and its phase transition behaviors but also lay the groundwork for applying machine learning methods to analyze complex physical systems.

\section{Conclusion}
In this paper, we investigated the tricritical behavior of the tricritical directed percolation model through a combination of Monte Carlo simulations and machine learning techniques. We first explored the system dynamics under various values of $q$ and $p$, using Monte Carlo simulations to identify the phase boundaries between absorbing and active phases. By analyzing the steady-state particle density, we determined the existence of both first-order and second-order phase transitions and identified the tricritical point, where these transitions meet.

Next, we applied a convolutional neural network to recognize phase transitions from the configuration data generated by simulations. The neural network successfully identified the critical points $p_c(q)$ for different values of $q$ without prior knowledge of the tricritical point. Additionally, by training the network on steady-state configurations at the upper phase boundary $p_u$, we identified the tricritical point at $q_t = 0.8953$, consistent with results obtained from Monte Carlo simulations.

Through the morphological analysis generated by Monte Carlo simulations, we found that the TCP model exhibits first-order and second-order phase transitions. Between these phase transition boundaries, there exist tricritical points and their crossover regions. We successfully identified the tricritical points and crossover regions using a three-output CNN model. This result is consistent with the findings from Monte Carlo simulations, yielding the phase boundary diagram shown in Fig~\ref{pq_ans}.

Our approach demonstrates the effectiveness of combining traditional simulation methods with modern machine learning techniques in studying phase transitions, especially in complex systems like the TDP model. The neural network method not only provided accurate identification of critical points but also revealed the tricritical point and its associated crossover region. This work provides a framework for future studies on phase transitions and critical phenomena using machine learning methods.
\section{Acknowledgements}
This work was supported in part by Yunnan Fundamental Research Projects (Grant 202401AU070035), Research Fund of Baoshan University(BYPY202216, BSKY202305), and the 111 Project 2.0, with Grant No. BP0820038.

\section{Data sets}
The detailed algorithms of how to generate raw data and implement machine learning are shown in the GitHub link {https://github.com/ChuckShen/perc-snn}.

\bibliographystyle{apsrev4-2}
\bibliography{ref}

\end{document}